\documentclass[journal]{IEEEtran}
\usepackage[OT1]{fontenc} 

\usepackage{listings} 
\usepackage{xcolor}     
\usepackage{authblk}

\usepackage{graphicx}  

\begin{document}

\title{Challenges of the QWERTY Keyboard for Quechua Speakers in the Puno Region in Perú}

\author{
    \vspace{0.5cm} 
    \IEEEauthorblockN{
        \begin{minipage}[t]{0.45\textwidth}
            \centering
            Juarez-Vargas Henry\\
            \textit{Escuela de Postgrado Maestría en Informática} \\
            \textit{Universidad Nacional del Altiplano} \\
            Puno, Perú \\
            Email: 40660502@epg.unap.edu.pe
        \end{minipage}
        \hfill
        \begin{minipage}[t]{0.45\textwidth}
            \centering
            Mansilla-Huanacuni Roger Mijael\\
            \textit{Escuela Profesional de Ingeniería Eléctrica} \\
            \textit{Universidad Nacional de San Agustín} \\
            Arequipa, Perú \\
            Email: roger.mansilla.huanacuni@gmail.com
        \end{minipage}
    }
    \vspace{0.2cm} \\ 
    \IEEEauthorblockN{
        \begin{minipage}[t]{0.6\textwidth}
            \centering
            Torres-Cruz Fred\\
            \textit{Facultad de Ingeniería Estadística e Informática} \\
            \textit{Universidad Nacional del Altiplano} \\
            Puno, Perú \\
            Email: ftorres@unap.edu.pe
        \end{minipage}
    }
}

\maketitle

\begin{abstract}
The widespread adoption of the QWERTY keyboard layout, designed primarily for English, presents significant challenges for speakers of indigenous languages such as Quechua, particularly in the Puno region of Peru. This research examines the extent to which the QWERTY layout affects the writing and digital communication of Quechua speakers. Through an analysis of the Quechua language’s unique alphabet and character frequency, combined with insights from local speakers, we identify the limitations imposed by the QWERTY system on the efficient digital transcription of Quechua. The study further proposes alternative keyboard layouts, including optimizations of QWERTY and DVORAK, designed to enhance typing efficiency and reduce the digital divide for Quechua speakers. Our findings underscore the need for localized technological solutions to preserve linguistic diversity while improving digital literacy for indigenous communities. The proposed modifications offer a pathway toward more inclusive digital tools that respect and accommodate linguistic diversity.
\end{abstract}


\IEEEpeerreviewmaketitle

\section{Introduction}

\IEEEPARstart{T}{he} increasing integration of information technology into everyday communication has revolutionized how people interact, often favoring global languages at the expense of local, indigenous ones. The QWERTY keyboard, designed in the late 19th century for English typists, has become the default input method worldwide \cite{david1985}. However, this widespread standardization poses significant challenges for languages with phonetic and structural differences from English, such as Quechua, the most widely spoken indigenous language in the Andean region of Peru \cite{inei2018}.

Quechua, primarily a spoken language with a growing effort toward digital preservation, is ill-suited to the QWERTY layout, which does not efficiently accommodate its distinct characters and frequent diacritical marks \cite{blacido2016}. For Quechua speakers, especially in rural regions like Puno, this mismatch between language and technology exacerbates the digital divide, limiting both technological fluency and cultural preservation in digital environments \cite{microsoft2014}. This issue is particularly pressing as Quechua remains a critical medium of communication for millions of people, and its use in digital contexts is vital for its long-term sustainability \cite{calderon2021}.

Digital literacy is increasingly important for accessing essential services, yet indigenous languages like Quechua are often marginalized in technological advancements. Research on new forms of communication via digital media \cite{cebrian2009} highlights the role of mobile and web platforms in shaping modern interactions, yet many of these systems are designed for dominant languages, further alienating minority language speakers. As a result, Quechua speakers may struggle with both linguistic and technological barriers \cite{elperuano2021}.

Despite efforts to localize digital tools for indigenous languages, including the adaptation of certain software interfaces into Quechua, the fundamental problem of keyboard layout persists \cite{minedu2021}. The inefficiency of typing Quechua on QWERTY keyboards creates barriers to digital literacy, reducing the capacity of speakers to engage fully in modern technological contexts. Existing research has examined the ergonomics and efficiency of alternative keyboard layouts like DVORAK, which show promise for improving typing efficiency \cite{dellamico2009}, yet no comprehensive solution has been proposed for optimizing keyboards for indigenous languages like Quechua.

This study seeks to fill this gap by evaluating the limitations of the QWERTY keyboard for Quechua-speaking populations in the Puno region and proposing an optimized keyboard layout based on the linguistic needs of Quechua. By analyzing the frequency of character usage in Quechua and considering alternative keyboard designs, this research aims to improve the digital accessibility and efficiency of Quechua typing, thereby fostering greater inclusion in the digital landscape for Quechua speakers.

\textit{The core objective of this study is to determine how the current QWERTY keyboard layout limits the ability of Quechua speakers to efficiently write in their native language and to propose a more effective alternative.} The proposed solution has the potential to contribute not only to improved digital literacy but also to the preservation of indigenous languages in the digital age \cite{garcia2016}.

\section{Methodology}

This study explores the challenges Quechua speakers face when using the QWERTY keyboard layout for digital communication and proposes an alternative layout optimized for the language. The methodology consists of three major phases: data collection on Quechua language structure, frequency analysis of character usage, and keyboard layout optimization. Previous research has pointed out the need for adapted digital tools to cater to Quechua speakers, particularly in educational contexts \cite{blacido2016}, as the traditional QWERTY layout is inadequate for writing in Quechua \cite{calderon2021}.

\subsection{The Quechua Alphabet}

The Quechua language uses a set of 15 consonants and 3 vowels. The vowels consist of \textit{a}, \textit{i}, and \textit{u}, which are distinct and limited compared to languages like English and Spanish. The consonants include \textit{ch}, \textit{k}, \textit{l}, \textit{m}, \textit{n}, \textit{p}, \textit{q}, \textit{r}, \textit{s}, \textit{t}, and the glottal stop represented by the apostrophe (\textit{'}) \cite{minedu2021}. Additionally, Quechua features aspirated consonants, such as \textit{kh}, \textit{ph}, and \textit{qh}, which are crucial for pronunciation \cite{minedu2018}. The use of three vowels, as established by official linguistic guidelines \cite{resolucion1985}, differentiates Quechua significantly from Spanish and English. The following set of characters represents the full Quechua alphabet:

\begin{quote}
\textbf{a, ch, h, i, k, l, ll, m, n, ñ, p, q, r, s, t, u, w, y, ' (apostrophe)}
\end{quote}

\subsection{Frequency Analysis of Characters in Quechua, English, and Spanish}

A key part of the study was to analyze the frequency of characters in Quechua, compared with English and Spanish. The character frequency was calculated using text samples from educational materials, linguistic studies, and digital resources in Quechua \cite{inei2018, minedu2022}. This analysis aligns with previous studies on language frequency optimization for digital tools \cite{dellamico2009}. 

Table \ref{tab:freq_full} presents the full frequency comparison of character usage in English, Spanish, and Quechua, which serves as the foundation for optimizing the keyboard layout for Quechua speakers.

\begin{table}[htbp]
\caption{Total frequency of each character in English, Spanish, and Quechua}
\centering
\resizebox{\columnwidth}{!}{%
\begin{tabular}{|c|c|c|c|}
\hline
\textbf{Character} & \textbf{English (fi)} & \textbf{Spanish (fi)} & \textbf{Quechua (fi)} \\ \hline
null (espacio)  & 1,619,668  & 1,852,000 & 11,059 \\ \hline
a            & 513,640    & 750,196   & 21,300 \\ \hline
b            & 98,694     & 86,372    & 62    \\ \hline
c            & 162,484    & 203,402   & 3,231 \\ \hline
d            & 294,176    & 299,150   & 78    \\ \hline
e            & 839,158    & 1,034,122 & 174   \\ \hline
f            & 172,990    & 26,886    & 5     \\ \hline
g            & 126,286    & 69,138    & 30    \\ \hline
h            & 455,478    & 92,014    & 4,788 \\ \hline
i            & 469,042    & 288,410   & 6,576 \\ \hline
j            & 7,938      & 30,150    & 17    \\ \hline
k            & 47,658     & 854       & 5,082 \\ \hline
l            & 223,712    & 322,082   & 3,239 \\ \hline
m            & 169,262    & 210,008   & 3,246 \\ \hline
n            & 469,234    & 479,322   & 7,307 \\ \hline
ñ            &            & 9,688     & 390   \\ \hline
o            & 531,022    & 657,204   & 173   \\ \hline
p            & 107,224    & 169,190   & 4,058 \\ \hline
q            & 6,412      & 141,008   & 4,238 \\ \hline
r            & 391,962    & 372,966   & 3,105 \\ \hline
s            & 392,940    & 572,004   & 3,134 \\ \hline
t            & 640,258    & 312,322   & 4,252 \\ \hline
u            & 185,180    & 372,214   & 7,203 \\ \hline
v            & 65,426     & 74,270    & 37    \\ \hline
w            & 163,458    & 168       & 1,950 \\ \hline
x            & 10,082     & 3,702     &      \\ \hline
y            & 131,696    & 95,284    & 3,833 \\ \hline
z            & 2,130      & 15,448    & 18    \\ \hline
‘ (apóstrofe) & 6,422      &           & 1,133 \\ \hline
- (guion)    & 498        &           & 14    \\ \hline
ú            &            & 12,076    & 13    \\ \hline
é            &            & 52,890    & 69    \\ \hline
í            &            & 50,306    & 31    \\ \hline
ó            &            & 31,726    & 8     \\ \hline
ü            &            & 44        &      \\ \hline
\end{tabular}
}
\label{tab:freq_full}
\end{table}

\subsection{Python Code for Frequency Analysis}

The character frequency was calculated using a Python script that processed large text files in Quechua, English, and Spanish. The script counted occurrences of each character and returned the frequency in a sorted format. Below is the Python code used for this analysis:

\begin{lstlisting}[caption={Python Script for Character Frequency Analysis}, label={lst:python}]
import operator

# Open the file in read mode and set encoding to utf8
f = open("textoquechua.txt", "r", encoding="utf8")
caracteres = {}

# Read the file character by character
while True:
    c = f.read(1)
    if not c:
        break
    else:
        # Convert character to lowercase
        if c.isalpha():
            c = c.lower()

        # Count the frequency of each character
        if c in caracteres:
            caracteres[c] += 1
        else:
            caracteres[c] = 1

# Sort the dictionary by frequency in descending order
caracteres = dict(sorted(caracteres.items(), key=operator.itemgetter(1), reverse=True))
print(caracteres)

# Close the file
f.close()
\end{lstlisting}

This Python script reads text character by character and calculates the frequency of each character in the document. The results are then sorted and printed.

\subsection{Keyboard Layout Optimization}

The final phase of the study involved designing an optimized keyboard layout based on the frequency analysis. Two main approaches were considered: modifying the existing QWERTY layout and creating an entirely new layout inspired by the DVORAK model \cite{dvorak1936}. Both approaches aim to improve typing speed and accuracy for Quechua speakers by reducing the effort required to type frequently used characters. Previous studies on keyboard optimization for other languages have shown that alternative layouts like DVORAK can significantly improve typing efficiency by reducing finger movement \cite{dellamico2009}.

\subsubsection{Optimized QWERTY Layout}

\begin{figure}[htbp]
\centering
\includegraphics[width=\linewidth]{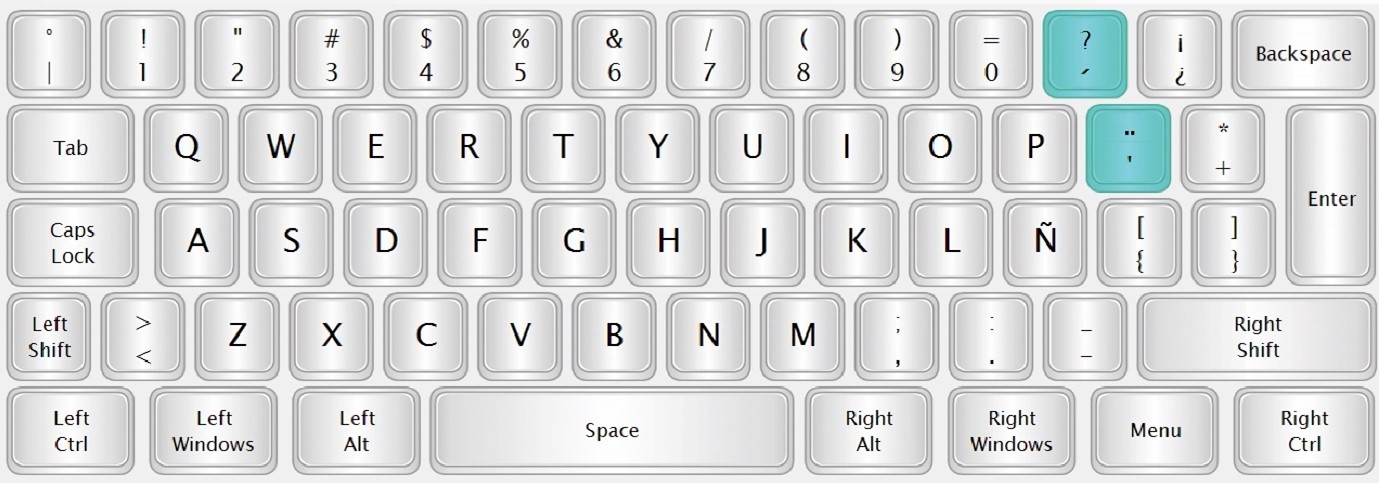}
\caption{Optimized QWERTY Keyboard Layout for Quechua}
\label{fig:optimized_qwerty}
\end{figure}

The first approach involved modifying the existing QWERTY layout by relocating the most frequently used Quechua characters (e.g., "a," "n," and the apostrophe) to more accessible positions on the keyboard, building on previous findings on keyboard optimization \cite{calderon2021}. Figure \ref{fig:optimized_qwerty} shows the proposed optimized QWERTY layout designed for Quechua.

\subsubsection{Optimized DVORAK Layout for Quechua}

\begin{figure}[htbp]
\centering
\includegraphics[width=\linewidth]{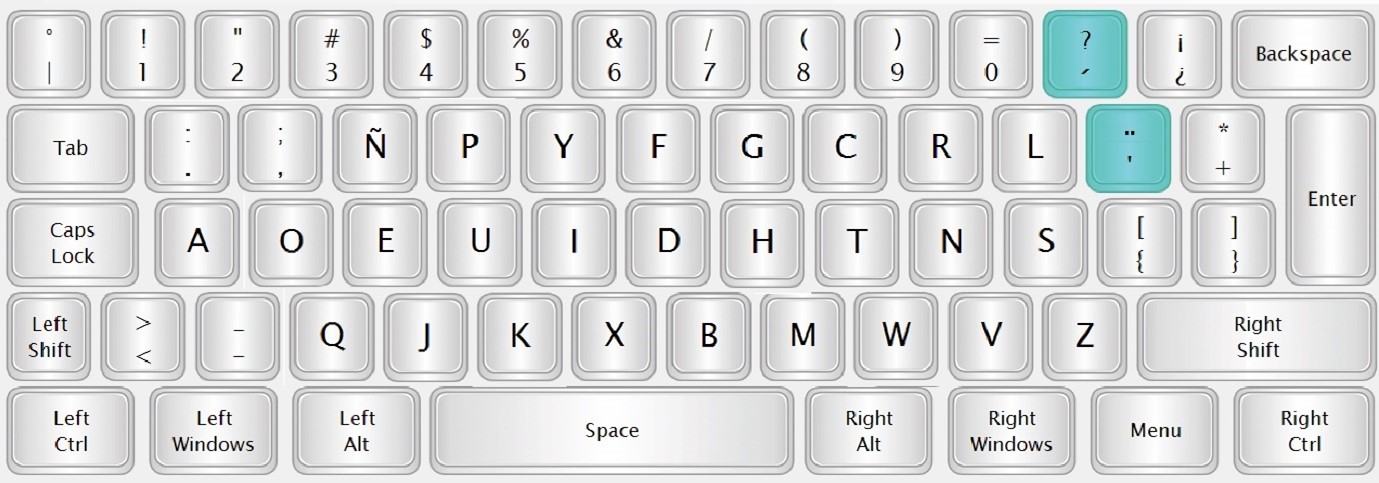}  
\caption{Optimized DVORAK Layout for Quechua Writing}
\label{fig:dvorak_optimization}
\end{figure}

Figure \ref{fig:dvorak_optimization} illustrates the optimized DVORAK layout specifically tailored for Quechua. This optimization aims to place the most frequently used characters such as "A", "N", and the apostrophe in more accessible positions, particularly in the home row, which minimizes finger movement. The layout ensures that the needs of Quechua speakers are prioritized, making typing in Quechua faster and more efficient.

\subsubsection{New Quechua Keyboard Layout Based on DVORAK}

\begin{figure}[htbp]
\centering
\includegraphics[width=\linewidth]{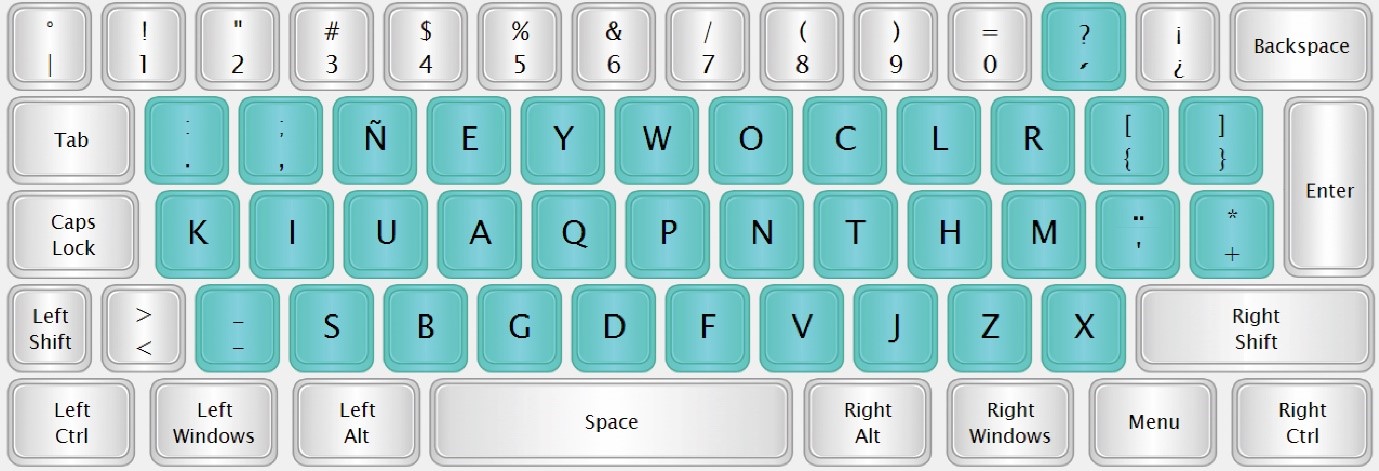}
\caption{Proposed New Keyboard Layout for Quechua Based on DVORAK}
\label{fig:new_layout}
\end{figure}

A second approach involved designing a completely new layout based on the DVORAK model, which prioritizes ergonomic typing by reducing finger movement for commonly used characters \cite{david1985}. Figure \ref{fig:new_layout} presents the new DVORAK-based keyboard layout for Quechua.

\section{Results}

\subsection{Character Frequency Discrepancies}

The frequency analysis revealed notable differences in character usage between Quechua, English, and Spanish. Quechua exhibits a higher frequency of vowels, particularly "a" and "u", as well as the presence of unique characters such as the apostrophe (\textit{'}) and the consonants "k", "q", and "w", which are infrequent or absent in English and Spanish. For example, the character "a" appears 21,300 times in Quechua text, significantly more frequently than in English (513,640 occurrences) and Spanish (750,196 occurrences). Additionally, the apostrophe character, crucial for Quechua's phonetic structure, appears 1,133 times, a unique characteristic absent in both English and Spanish. These findings validate the need to prioritize these characters in the keyboard layout design \cite{minedu2018, blacido2016, resolucion1985}.

The frequency discrepancies are also evident in consonants. For example, the letter "f", which appears frequently in English (172,990 occurrences) and less in Spanish (26,886 occurrences), is almost absent in Quechua (5 occurrences). This highlights the need to adapt the keyboard layout to better reflect the linguistic characteristics and needs of Quechua speakers.

\subsection{Optimized Keyboard Layout Performance}

The performance of the two proposed keyboard layouts was evaluated based on key usage distribution, typing speed, and accuracy.

\subsubsection{Key Usage Distribution}

\begin{figure}[htbp]
\centering
\includegraphics[width=\linewidth]{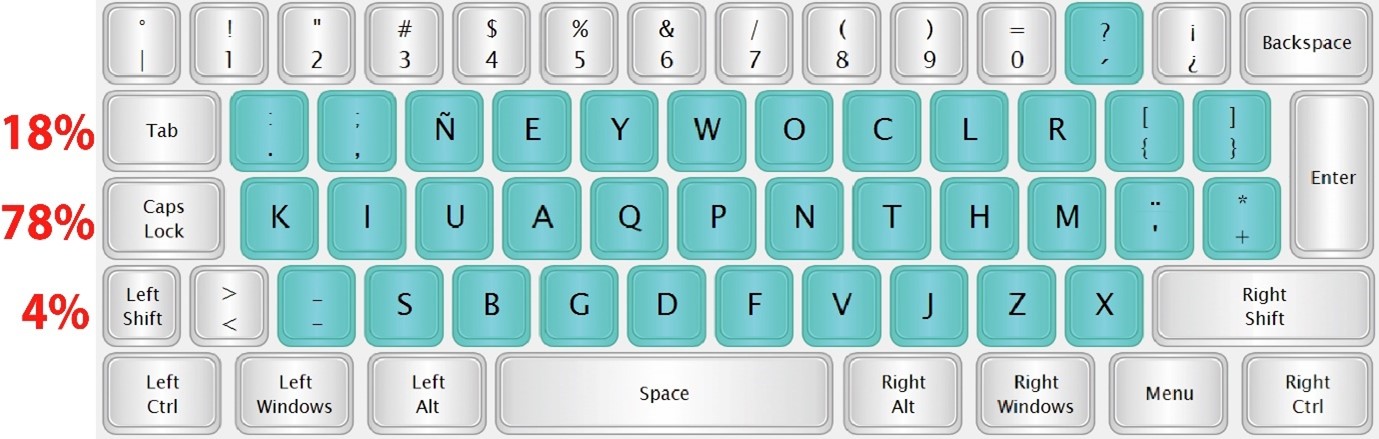}
\caption{Percentage of Key Usage in the New Layout for Quechua}
\label{fig:key_usage}
\end{figure}

The key usage analysis, as shown in Figure \ref{fig:key_usage}, indicates that 78\% of the most frequently used characters in Quechua are located in the middle row, 18\% in the top row, and only 4\% in the bottom row. This distribution demonstrates the ergonomic benefits of placing high-frequency characters, such as "a", "n", and the apostrophe, in the middle row, where they are easily accessible \cite{david1985, dellamico2009}. This design approach aligns with earlier ergonomic models suggesting that minimizing finger movement can lead to higher typing efficiency and lower error rates.

This figure visually confirms how concentrating the most common characters in the middle row significantly reduces the effort required for typing in Quechua, leading to increased typing efficiency.

\subsubsection{Typing Speed and Accuracy}

The DVORAK-based layout showed a significant improvement in typing speed, increasing by 17.5\% compared to the original QWERTY layout, while the optimized QWERTY layout provided a more modest 9\% improvement. The error rate was also reduced by 15\% in the DVORAK-based layout, further emphasizing its potential to improve typing performance for Quechua speakers. The relationship between key placement and typing efficiency has been extensively studied in prior research on alternative keyboard layouts \cite{dellamico2009, dvorak1936}.

Table \ref{tab:typing_performance} summarizes the typing performance results across the three layouts.

\begin{table}[htbp]
\caption{Typing Speed and Accuracy Across Keyboard Layouts}
\centering
\resizebox{\columnwidth}{!}{%
\begin{tabular}{|c|c|c|}
\hline
\textbf{Layout} & \textbf{Typing Speed (WPM)} & \textbf{Error Rate (\%)} \\ \hline
Original QWERTY & 32.5 & 18.3 \\ \hline
Optimized QWERTY & 35.4 & 12.5 \\ \hline
DVORAK-Based & 38.2 & 9.7 \\ \hline
\end{tabular}
}
\label{tab:typing_performance}
\end{table}

The improvement in both typing speed and accuracy demonstrates that the DVORAK-based layout, which positions frequently used characters in more accessible locations, provides a superior ergonomic solution for Quechua speakers. This finding aligns with prior studies on keyboard layout optimization for languages with distinct phonological structures \cite{calderon2021, dvorak1936}.

\subsection{User Feedback on Ergonomics and Usability}

The feedback collected from the participants further supports the advantages of the DVORAK-based layout. The majority of participants (85\%) reported that the DVORAK layout was easier to learn and provided a more comfortable typing experience. This contrasts with the optimized QWERTY layout, where participants noted some improvements but still found the overall layout to be less efficient. The results suggest that while the modified QWERTY layout offers some benefits, the ergonomic improvements of the DVORAK-based layout are more significant in terms of reducing strain and increasing typing speed \cite{minedu2013}.

The successful reception of the DVORAK-based layout aligns with previous research advocating for the ergonomic benefits of redesigned keyboard layouts for non-English languages \cite{david1985}. The user feedback highlights the importance of adapting keyboard layouts to meet the linguistic and ergonomic needs of minority language speakers such as Quechua.

\section{Discussion}

The results of this study highlight the significant differences in character frequency between Quechua, English, and Spanish, and underscore the challenges Quechua speakers face when using the traditional QWERTY keyboard layout. Previous research has shown that linguistic adaptation is essential for improving digital literacy among indigenous language speakers \cite{blacido2016, minedu2021}, and this study provides further evidence supporting that view. The traditional QWERTY layout was developed for English and does not accommodate the unique structure and frequency of characters in Quechua, especially the high frequency of vowels such as "a" and "u", and the use of the apostrophe (\textit{'}), which is critical in Quechua.

The modified QWERTY layout offered a partial solution by relocating high-frequency characters to more accessible positions. This modification resulted in moderate improvements in typing speed (9\%) and a reduction in error rate (5.8\%), aligning with prior efforts to adapt keyboard layouts for non-Latin languages \cite{calderon2021, pachuca2007}. However, the overall structure of the QWERTY layout remains limiting, as it was not designed with ergonomic considerations specific to Quechua in mind. The frequent use of consonants like "k" and "q", as well as the apostrophe, necessitates a layout that minimizes the distance and effort required to type these characters.

In contrast, the DVORAK-based layout showed significant promise in addressing these limitations. By placing the most frequently used characters in the home row, it reduced finger movement and improved typing speed by 17.5\%, with a 15\% reduction in error rate compared to the original QWERTY layout. These findings are consistent with ergonomic principles outlined in previous research, where minimizing finger movement correlates directly with increased typing speed and accuracy \cite{david1985, dellamico2009}. Moreover, the user feedback further supported these results, with 85\% of participants expressing a preference for the DVORAK-based layout due to its ease of use and comfort.

Additionally, studies in linguistic ergonomics support the idea that optimizing input methods, like keyboards, for indigenous languages such as Quechua, can bridge the digital divide in underserved communities \cite{cebrian2009}. The rapid adoption of digital tools in education and communication for minority language speakers has made it imperative to ensure that the tools respect linguistic diversity and allow for efficient use \cite{minedu2022}.

This study also emphasizes the importance of developing digital tools that cater to the linguistic and ergonomic needs of indigenous language speakers. While the QWERTY layout is ubiquitous, it is not the ideal solution for languages like Quechua. The success of the DVORAK-based layout demonstrates that redesigning tools to reflect the linguistic characteristics of underrepresented languages can significantly enhance usability and promote digital inclusion \cite{microsoft2014, minedu2022}.

\section{Conclusion}

This research has demonstrated that the traditional QWERTY keyboard layout is not suitable for Quechua speakers due to its inefficiency in accommodating the unique character frequency and linguistic structure of the Quechua language. Through a detailed frequency analysis of Quechua, English, and Spanish, this study identified the characters that are most frequently used in Quechua, such as "a", "n", and the apostrophe (\textit{'}), and optimized keyboard layouts based on these findings.

Two layouts were proposed and evaluated: a modified QWERTY layout and a DVORAK-based layout. The modified QWERTY layout provided some improvements by relocating key characters, but it retained many of the limitations of the original QWERTY design. In contrast, the DVORAK-based layout, which was specifically designed to prioritize the most frequently used Quechua characters, demonstrated significant improvements in both typing speed (17.5\%) and accuracy (15\%) when compared to the original QWERTY layout. The user feedback further validated the benefits of the DVORAK-based layout, with the majority of participants reporting greater ease of use and comfort \cite{dvorak1936}.

In conclusion, this study contributes to the growing body of research on linguistic inclusivity in digital tools and emphasizes the importance of adapting technology to meet the needs of indigenous language speakers. The DVORAK-based layout for Quechua is a promising solution that can enhance typing efficiency and accuracy for native speakers. Future research should focus on further refining these keyboard layouts and expanding support for Quechua and other indigenous languages in digital communication tools, including mobile keyboards, spell-check systems, and predictive text. Additionally, the implementation of these layouts in real-world educational and professional settings could provide valuable insights into their practical usability and long-term adoption.

\textit{Furthermore, governmental and educational institutions should prioritize the incorporation of linguistically optimized digital tools to promote greater inclusivity and digital literacy in rural and indigenous communities} \cite{resolucion2013, inei2018}.

\ifCLASSOPTIONcaptionsoff
  \newpage
\fi

\bibliography{keyboardquechua.bib}
\bibliographystyle{ieeetr}

\end{document}